\documentclass[10pt,aps,prb,longbibliography,twocolumn,preprintnumbers,amsmath,amssymb,floatfix,showpacs,citeautoscript]{revtex4-2}
\usepackage[latin1]{inputenc}
\usepackage[T1]{fontenc}
\usepackage[english]{babel}
\usepackage[pdftex]{graphicx}
\usepackage{amsmath}
\usepackage{times}
\usepackage[usenames,dvipsnames,svgnames,table]{xcolor}
\usepackage[colorlinks,plainpages=false,linkcolor=blue,urlcolor=blue,citecolor=blue,pdfpagemode=UseNone]{hyperref}

\renewcommand{\AA}{\text{\r{A}}}

\newcommand\Multp{\cdot}
\newcommand\Vek[1]{\vec{#1}}

\newcommand\lc[1]{\lowercase{#1}}

\begin{document}

\title
{
\boldmath
Correlated interface electron gas in infinite-layer nickelate versus cuprate films on SrTiO$_3$(001)
}

\author{Benjamin Geisler}
\email{benjamin.geisler@uni-due.de}
\affiliation{Department of Physics and Center for Nanointegration (CENIDE), Universit\"at Duisburg-Essen, Lotharstr.~1, 47057 Duisburg, Germany}
\author{Rossitza Pentcheva}
\email{rossitza.pentcheva@uni-due.de}
\affiliation{Department of Physics and Center for Nanointegration (CENIDE), Universit\"at Duisburg-Essen, Lotharstr.~1, 47057 Duisburg, Germany}

\date{\today}

\begin{abstract}
Based on first-principles calculations including a Coulomb repulsion term, we identify trends in the electronic reconstruction 
of $A$NiO$_2$/SrTiO$_3$(001) ($A=$~Pr, La) and $A$CuO$_2$/SrTiO$_3$(001) ($A=$~Ca, Sr).
Common to all cases is the emergence of a quasi-two-dimensional electron gas (q2DEG) in SrTiO$_3$(001), 
albeit the higher polarity mismatch at the interface of nickelates vs.\ cuprates to the nonpolar SrTiO$_3(001)$ substrate
(${3+}/0$ vs.\ ${2+}/0$)
results in an enhanced q2DEG carrier density.
The simulations reveal a significant dependence of the interfacial Ti $3d_{xy}$ band bending on the rare-earth ion in the nickelate films,
being $20$-$30\%$ larger for PrNiO$_2$ and NdNiO$_2$ than for LaNiO$_2$.
Contrary to expectations from the formal polarity mismatch,
the electrostatic doping in the films is twice as strong in cuprates as in nickelates.
We demonstrate that the depletion of the self-doping rare-earth $5d$ states enhances the similarity of nickelate and cuprate Fermi surfaces in film geometry,
reflecting a single hole in the Ni and Cu $3d_{x^2-y^2}$ orbitals.
Finally, we show that NdNiO$_2$ films grown on a polar NdGaO$_3(001)$ substrate feature a simultaneous suppression of q2DEG formation as well as Nd~$5d$ self-doping.
\end{abstract}


\maketitle

\section{Introduction}

The very recent observation of superconductivity in Sr-doped NdNiO$_2$ and PrNiO$_2$ films grown on SrTiO$_3$(001) (STO)
\cite{Li-Supercond-Inf-NNO-STO:19, Li-Supercond-Dome-Inf-NNO-STO:20, Osada-PrNiO2-SC:20}
has sparked considerable interest in infinite-layer nickelates,
since their formal Ni$^{1+}$ ($3d^9$) valence state renders them close to cuprates~\cite{Nomura-Inf-NNO:19, JiangZhong-InfNickelates:19, Sawatzky-NNO:19, Sakakibara:19, JiangBerciuSawatzky:19, Botana-Inf-Nickelates:19, Kitatani-AritaZhongHeld:20, Si-Zhonh-Held:InfNNO-Hydrogen:20, ChoiMiYoungPickett:20, Lechermann-Inf:20, NNO-SelfDopingDesign-d9-Arita:20, NNO-SC-Thomale:20, BernardiniCano:20}.
In the quest for a fundamental understanding of the underlying mechanisms, several aspects are noteworthy and so far unresolved.
In particular, superconductivity could not be confirmed experimentally in Sr-doped bulk NdNiO$_2$~\cite{Li-NoSCinBulkDopedNNO:19}
and was not observed in LaNiO$_2$ films on STO(001)~\cite{Li-Supercond-Inf-NNO-STO:19} despite its similar electronic structure to NdNiO$_2$ and PrNiO$_2$ in the bulk, apart from the Nd and Pr~$4f$ states~\cite{Choi-Lee-Pickett-4fNNO:20}.

A considerable electronic reconstruction emerges in NdNiO$_2$/SrTiO$_3$(001)
due to the polar discontinuities at the interface and the surface~\cite{GeislerPentcheva-InfNNO:20},
which comprises
(i)~the formation of a quasi-two-dimensional electron gas (q2DEG) at the interface by occupation of Ti $3d$ states despite the metallic screening of the nickelate film; (ii)~the depletion of the self-doping Nd $5d$ states, resulting in a cuprate-like Fermi surface; and (iii)~an enhanced and modulated Ni $e_g$ orbital polarization throughout the film due to electrostatic doping.
Notably, the q2DEG was found to be far more pronounced than its counterpart emerging
in the paradigmatic LaAlO$_3$/SrTiO$_3$(001) system (LAO/STO~\cite{Ohtomo:2004, Nakagawa:06})
beyond 4 monolayers (ML) of LAO~\cite{Thiel:06, PentchevaPickett:09},
which is known to exhibit intriguing correlation-driven physics such as superconductivity~\cite{Reyren:07}.

In this context, the analogies frequently drawn between infinite-layer nickelates and cuprates~\cite{JiangZhong-InfNickelates:19, Botana-Inf-Nickelates:19, Lechermann-Inf:20} appear in a novel light.
A fundamental difference between these two materials classes remains, namely, the formal polarity of the consecutive (001) layers,
being, for instance,
Ca$^{2+}$(CuO$_2$)$^{2-}$ and Sr$^{2+}$(CuO$_2$)$^{2-}$
in the superconducting infinite-layer cuprates
and
Nd$^{3+}$(NiO$_2$)$^{3-}$, Pr$^{3+}$(NiO$_2$)$^{3-}$, and La$^{3+}$(NiO$_2$)$^{3-}$
in the infinite-layer nickelates.
This implies distinct behavior between cuprates and nickelates in film geometry on a nonpolar substrate such as STO(001)
despite the same formal $3d^9$ configuration.

Here we systematically explore the impact of the polar discontinuities
at the interface and the surface
on the structural and electronic properties
of PrNiO$_2$/STO(001), LaNiO$_2$/STO(001), CaCuO$_2$/STO(001), and SrCuO$_2$/STO(001)
in film geometry
by performing first-principles calculations including a Coulomb repulsion term.
In each system, we find that the polarity mismatch drives an electronic reconstruction,
inducing the formation of a q2DEG at the interface.
This supports the earlier finding of a q2DEG at the NdNiO$_2$/STO$(001)$ interface,
which is absent for perovskite films~\cite{GeislerPentcheva-InfNNO:20},
and establishes it as a general phenomenon in infinite-layer nickelate and cuprate films on STO$(001)$.
The occupation of the Ti $3d$ states varies in each case,
which reflects the different polar discontinuities of infinite-layer cuprates vs.\ nickelates at the interface to STO$(001)$,
and furthermore unravels substantial distinctions between NdNiO$_2$ and PrNiO$_2$ vs.\ LaNiO$_2$ films.
The electronic reconstruction is accompanied by ionic relaxations, specifically ferroelectric-like displacements of the Ti ions in the topmost $20$--$30~\AA$ of the STO substrate,
that act as a fingerprint of the q2DEG formation.
We demonstrate explicitly that the depletion of the rare-earth $5d$ states, which self-dope the bulk infinite-layer nickelates,
enhances the similarity of nickelate and cuprate Fermi surfaces in film geometry,
resulting in a single hole in the Ni and Cu $3d_{x^2-y^2}$ orbitals.
The hole density increases from the interface to the surface due to electrostatic doping, which we find to be twice as strong in cuprates as in nickelates,
contrary to expectations from the formal polarity mismatch.
Finally, we show that NdNiO$_2$ films grown on a polar NdGaO$_3(001)$ substrate exhibit depleted Nd~$5d$ states in the film
and a simultaneously quenched q2DEG in the substrate,
which offers a route to disentangle their contributions to superconductivity in infinite-layer nickelates.

\section{Methodology}

We performed first-principles calculations in the framework
of density functional theory~\cite{KoSh65} (DFT)
as implemented in the Quantum ESPRESSO code~\cite{PWSCF}.
The generalized gradient approximation was used for the exchange-correlation functional  
as parametrized by Perdew, Burke, and Ernzerhof (PBE)~\cite{PeBu96}.
Where indicated, we compare with PBEsol results, a functional that often renders improved structural properties~\cite{PBEsol:08, VermaGeislerPentcheva:19, Q2D:12}.
Higher-level methods are beyond the scope of the present work due to the large system sizes~\cite{Q2D:12, Q2D:16}.
Static correlation effects were considered within the DFT$+U$ formalism~\cite{Anisimov:93, QE-LDA-U:05}
employing $U=4$~eV on Ni, Cu, and Ti sites,
in line with previous work~\cite{Liu-NNO:13, Botana-Inf-Nickelates:19, Geisler-LNOSTO:17, WrobelGeisler:18, GeislerPentcheva-LNOLAO:18, GeislerPentcheva-LNOLAO-Resonances:19}.
We confirmed that a higher value of $U_\text{Cu} = 6.5$~eV~\cite{ZhongKosterKelly:12} leads to largely identical results.

We model $AB$O$_2$/STO$(001)$ in film geometry ($A=$~Pr, La, Ca, Sr; $B=$~Ni, Cu) by using $\sqrt{2}a \times \sqrt{2}a$ supercells with two transition metal sites per layer to account for octahedral rotations,
strained to the STO substrate lattice parameter $a = 3.905~\AA$.
The symmetric slabs contain $10.5$~ML of STO substrate and $4$~ML of infinite-layer nickelate or cuprate film on each side (the figures only show half of the supercell).
The vacuum region spans $20~\AA$.
Simulations using a NdGaO$_3(001)$ substrate are carried out in analogy ($a=3.86~\AA$).
%
%
Wave functions and density were expanded into plane waves up to cutoff energies of $45$ and $350$~Ry, respectively.
Ultrasoft pseudopotentials~\cite{Vanderbilt:1990}
as successfully employed in previous work~\cite{GeislerPopescu:14, Geisler-Heusler:15, GeislerFePcHSi:19, Geisler-LNOSTO:17, WrobelGeisler:18, GeislerPentcheva-LNOLAO:18, GeislerPentcheva-LNOLAO-Resonances:19, Viewpoint:19, GeislerPentcheva-LCO:20},
were used in conjunction with projector augmented wave datasets~\cite{PAW:94}.
The Pr and Nd $4f$ electrons are frozen in the core, similar to previous studies involving Nd~\cite{Liu-NNO:13, Nomura-Inf-NNO:19, Lechermann-Inf:20, GeislerPentcheva-InfNNO:20};
their explicit treatment leads to qualitatively similar results.
We used a $12 \times 12 \times 1$ Monkhorst-Pack $\Vek{k}$-point grid~\cite{MoPa76}
and $5$~mRy Methfessel-Paxton smearing~\cite{MePa89} to sample the Brillouin zone.
The ionic positions were accurately optimized, reducing ionic forces below $1$~mRy$/$a.u.

\begin{table}[b]
	\centering
	\vspace{-1.5ex}
	\caption{\label{tab:StructuralAspects}Structural data of the considered $AB$O$_2$/STO$(001)$ systems, compared to NdNiO$_2$/STO$(001)$ \cite{GeislerPentcheva-InfNNO:20}. The film thickness $d_{\text{Film}}$ is measured from the $B$ site positions in $S$ and $S-4$, spanning 4 ML [cf.~Fig.~\ref{fig:Structures}(a)]. The parameters $\tilde{z}$ and $\tilde{d}$ refer to the exponential fit of the ferroelectric-like Ti displacements in the STO substrate [cf.~Fig.~\ref{fig:Structures}(c)], as described in the text, and quantify how strong and deep the electronic reconstruction affects the substrate.}
	\begin{ruledtabular}
	\begin{tabular}{lccccc}
		 & NdNiO$_2$ & PrNiO$_2$ & LaNiO$_2$ & CaCuO$_2$ & SrCuO$_2$ \\
		\hline
		Strain at $a_\text{STO}$ (\%)	& $-0.4$ & $-0.4$ & $-1.4$ & $1.4$ & $-0.6$ \\
		$d_{\text{Film}}$ ($\AA$)	& $13.50$ & $13.68$ & $14.08$ & $13.44$ & $14.63$ \\
		$\tilde{z}$ ($\AA$)	& $-0.35$ & $-0.35$ & $-0.34$ & $-0.28$ & $-0.32$ \\
		$\tilde{d}$ ($\AA$)	& $7.42$ & $7.48$ & $7.66$ & $8.23$ & $8.05$ \\
	\end{tabular}
	\end{ruledtabular}
\end{table}

\section{Ionic response to the interface polarity}

\begin{figure}
	\centering
	\includegraphics[]{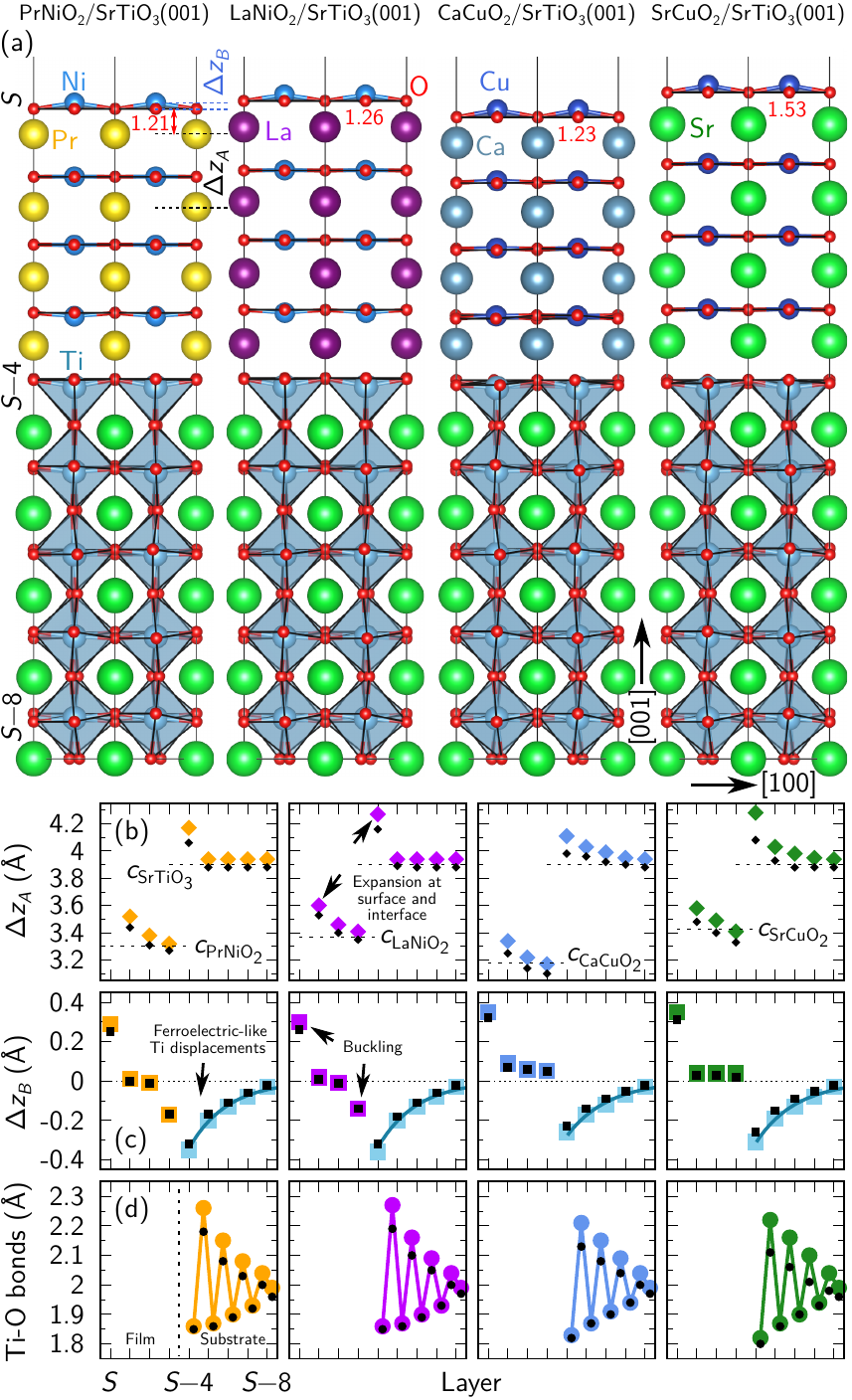}
	\caption{(a)~Optimized geometry of different $3d^9$ infinite-layer $AB$O$_2$/STO$(001)$ systems. The small red numbers denote the distance between the surface $B$O$_2$ layer and the subsurface $A$ layer. (b)~The apical $A$-site distances $\Delta z_A$ increase in the infinite-layer films from the interface to the surface. They are particularly enhanced at the interface ($S-4$), exceeding the STO bulk distance. The small horizontal dashed lines indicate bulk $c$ reference values. (c)~The $B$-O$_2$ displacements $\Delta z_B = z_B-z_\text{O}$ reveal a surface buckling in each case, whereas buckling at the interface (in the opposite direction) occurs only for the two nickelate systems. In the STO substrate, dark-blue curves represent a fit to an exponential function (see text and Table~\ref{tab:StructuralAspects}). (d)~The oscillating apical Ti-O bond lengths are linked to the $\Delta z_\text{Ti}$ displacements and act as a fingerprint of the q2DEG formation. --- In all panels, large colored symbols indicate PBE results, whereas small black symbols correspond to PBEsol values for comparison. \vspace{-1.5ex}}
	\label{fig:Structures}
\end{figure}

\begin{figure*}
	\centering
	\includegraphics[width=11cm]{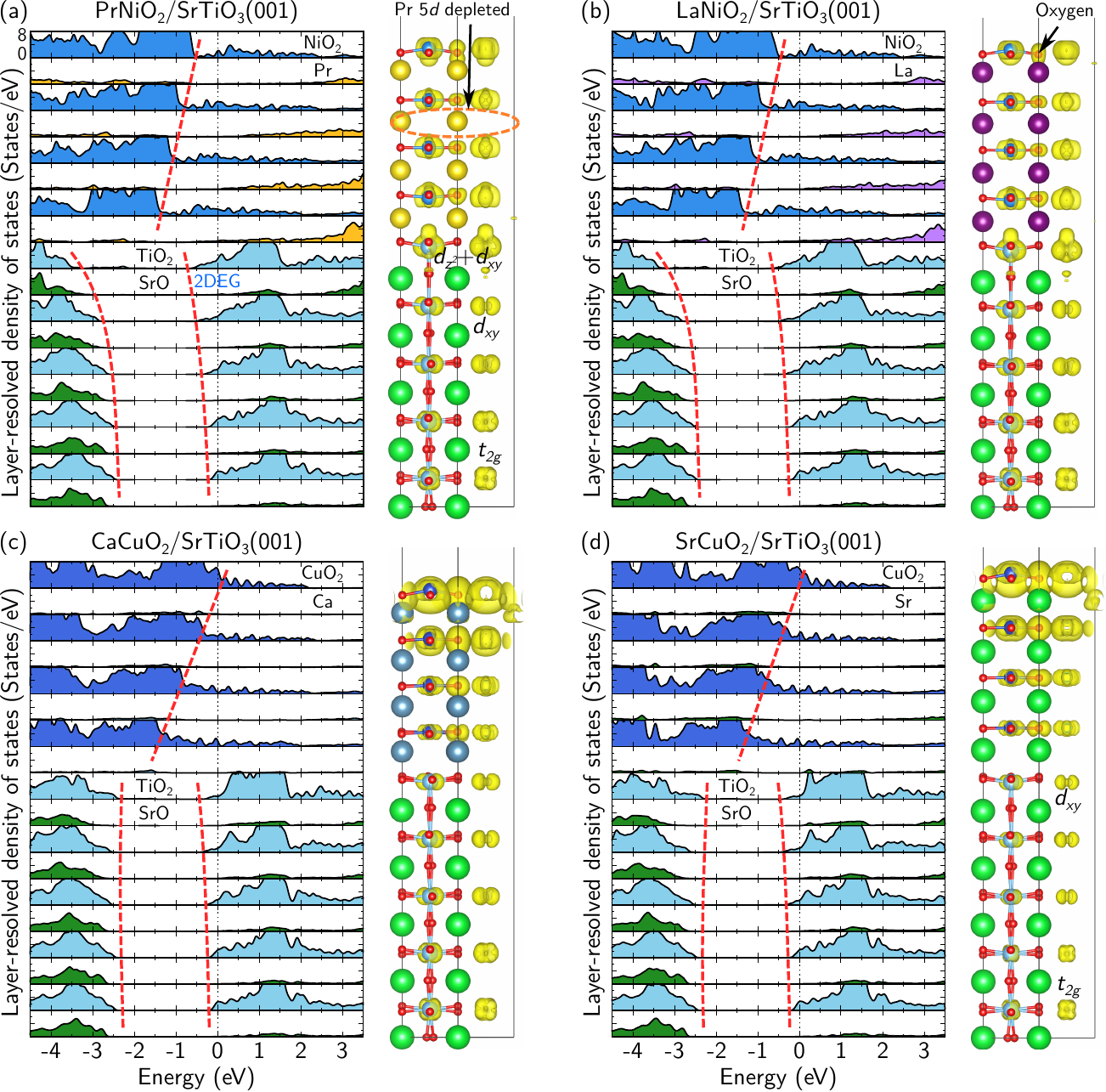}
	\caption{The layer-resolved densities of states of the different $AB$O$_2$/STO$(001)$ systems (cf.~Fig.~\ref{fig:Structures}) show the q2DEG formation at the interface due to the occupation of Ti $3d$ conduction band states. The band bending in the substrate due to polarity mismatch is more pronounced for nickelate films than for cuprate films (particularly for the valence band), whereas the emergent electric field in the films is approximately twice as strong in the cuprate films than in the nickelate films (as schematically indicated by red dashed lines; cf.~Fig.~\ref{fig:Bands}). The distribution of the electron density (integrated from $-0.7$~eV to $E_\text{F}$) visualizes the occupation of Ti $3d$ states that varies with the film composition (being less pronounced for cuprate films), the orbital order at the Ti sites, and the presence (absence) of a hybrid $d_{z^2}$ interface state for nickelate (cuprate) films. The absence of electron density in the rare-earth layers (at the rare-earth sites and the corresponding apical oxygen vacancy sites) highlights the 2D cuprate-like electronic structure emerging in the nickelate films.}
	\label{fig:LDOS-ILDOS}
\end{figure*}

The optimized geometries of different 4-ML $AB$O$_2$ films on STO(001)
are displayed in Fig.~\ref{fig:Structures}(a).
Similar to the case of NdNiO$_2$ ($a = 3.92$, $c = 3.28~\AA$~\cite{Hayward:03, Nomura-Inf-NNO:19}),
the lattice parameters of bulk PrNiO$_2$ ($a = 3.92$, $c = 3.30~\AA$; DFT+$U$), LaNiO$_2$ ($a = 3.96$, $c = 3.37~\AA$~\cite{Hayward:99, Botana-Inf-Nickelates:19}), and SrCuO$_2$ ($a = 3.93$, $c = 3.43~\AA$~\cite{Kobayashi:97, ZhongKosterKelly:12}) imply that these materials are subject to compressive strain, if grown epitaxially on STO(001) ($a = 3.905~\AA$),
whereas CaCuO$_2$ ($a = 3.85$, $c = 3.18~\AA$~\cite{Kobayashi:97, ZhongKosterKelly:12})
experiences tensile strain (Table~\ref{tab:StructuralAspects}).
Nevertheless, we observe a vertical expansion in all films,
as reflected in the apical $A$-site distances $\Delta z_A$ shown in Fig.~\ref{fig:Structures}(b).
This expansion is not uniform, but increases continuously from the interface to the surface.
In particular directly at the interface, the distances are enhanced;
this result can be associated with the electrostatic doping due to the polarity of the films,
similar to the previously observed enhanced La-Sr distance across
the $n$-type LaNiO$_3$/STO(001) interface ($\sim\!4.06~\AA$)~\cite{Geisler-LNOSTO:17, ZhangKeimer:14, Hwang:13}.
The SrCuO$_2$ case highlights that this expansion at the interface
is not exclusively related to a chemical variation at the $A$ site;
surprisingly, the effect is even strongest in this system.
Notably, for cuprate films this $\Delta z_A$ expansion extends several layers into the substrate,
which is clearly not the case for the nickelate systems
that show an abrupt transition to bulklike apical Sr-Sr distances in STO.
%
We find that PBE and PBEsol provide qualitatively similar structural properties (Fig.~\ref{fig:Structures}).
The small intrinsic octahedral rotations of STO are removed near the interface,
with the exception of the CaCuO$_2$ film,
and the $B$O$_4$ squares in the infinite-layer films show almost no rotations around the $c$ axis [Fig.~\ref{fig:Structures}(a)].

\begin{figure*}[t!]
	\centering
	\includegraphics[width=17cm]{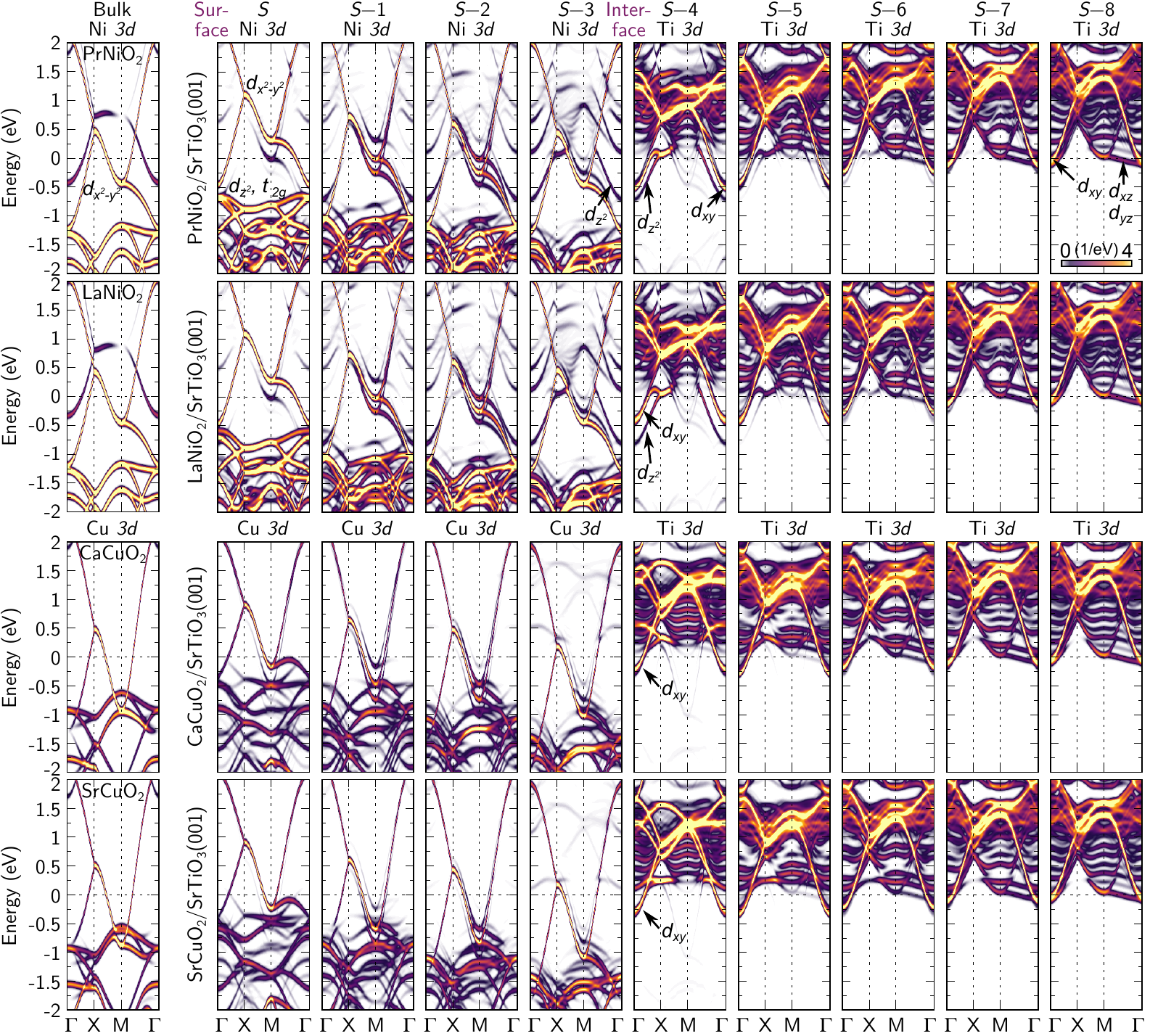}
	\caption{Band structure [$\Vek{k}$-resolved densities of states, projected on Ni, Cu, and Ti $3d$ orbitals in different layers (from left to right)] of the considered $AB$O$_2$/STO$(001)$ systems (cf.~Fig.~\ref{fig:Structures}). Corresponding bulk panels are provided for comparison. The orbital characters are denoted. The figure highlights the emergent q2DEG due to occupation of dispersive interfacial Ti $3d$ states in the STO substrate (predominantly $3d_{xy}$ states) and the modulation of the Ni/Cu $3d_{x^2-y^2}$ states throughout the infinite-layer films. Exclusively for the nickelate films, an interface state can be observed that exhibits a hybrid rare-earth $5d_{z^2}$--Ni $3d_{z^2}$--Ti $3d_{z^2}$ character.}
	\label{fig:Bands}
\end{figure*}

The cation-anion $B$-O$_2$ displacements $\Delta z_B = z_B-z_\text{O}$ shown in Fig.~\ref{fig:Structures}(c)
reveal a surface buckling in each case, whereas buckling at the interface (in the opposite direction) occurs exclusively for the nickelate systems.
This indicates that the bucking is primarily impacted by the $B$-site element.
In contrast, the distance between the surface $B$O$_2$ layer and the subsurface $A$ layer,
which is considerably contracted in each case,
correlates also with the ionic radius of the $A$ site element [Fig.~\ref{fig:Structures}(a)].

In the STO substrate, substantial ferroelectric-like displacements $\Delta z_B$ arise that are qualitatively similar for all considered systems [Fig.~\ref{fig:Structures}(c)].
For NdNiO$_2$/STO(001), such displacements were shown to be indicative of the q2DEG formation~\cite{GeislerPentcheva-InfNNO:20}.
A fit to $\Delta z_B = \tilde{z} \Multp \exp (- d / \tilde{d} )$ renders
the values given in Table~\ref{tab:StructuralAspects}.
Here, $\tilde{d}$ provides insight how deep the electronic reconstruction influences the ionic geometry in the STO substrate,
which is larger for cuprate films than for nickelate films.
In contrast, the maximal displacement $\tilde{z}$ is $\sim 0.05~\AA$ larger in the nickelate cases.
From these results, we estimate experimentally resolvable displacements (i.e., $0.35~\AA > \Delta z_B > 0.01~\AA$) in the topmost $20$--$30~\AA$ of the STO substrate.
The finite displacements $\Delta z_B$ are also reflected in the disproportionation of the apical Ti-O bond lengths, which oscillate strongly around the bulk value ($1.96~\AA$) in Fig.~\ref{fig:Structures}(d).

\begin{table*}[t]
	\centering
	\vspace{-1.5ex}
	\caption{\label{tab:BandBending}Band bending in the different $AB$O$_2$/STO$(001)$ systems. The band energies $\epsilon$ are given relative to the Fermi energy and refer to the $\Gamma$ point, where they are minimal (cf.~Fig.~\ref{fig:Bands}). $\Delta \epsilon$ denotes the band bending experienced by the planar $3d_{x^2-y^2}$ orbitals throughout the film from the interface ($S-3$) to the surface ($S$). We divide by the film thickness $d_{\text{Film}}$ (cf.~Table~\ref{tab:StructuralAspects}) to normalize. The interfacial Ti $3d_{xy}$ energies ($S-4$) reflect how pronounced the emerging q2DEG is. The presence of a partially occupied rare-earth $5d_{z^2}$--Ni $3d_{z^2}$ hybrid state in the bulk (the electron pocket at the $\Gamma$ point), admixed with Ti $3d_{z^2}$ in film geometry, distinguishes nickelates from cuprates (cf.~Figs.~\ref{fig:Bands} and~\ref{fig:NdLa5d}).}
	\begin{ruledtabular}
	\begin{tabular}{lccccc}
		 & NdNiO$_2$ \cite{GeislerPentcheva-InfNNO:20} & PrNiO$_2$ & LaNiO$_2$ & CaCuO$_2$ & SrCuO$_2$ \\
		\hline
		$\epsilon_{3d_{x^2-y^2}}^\text{Ni/Cu}$ in the bulk (eV)	& $-1.22$ & $-1.21$ & $-1.19$ & $-1.31$ & $-0.92$ \\
		$\epsilon_{3d_{x^2-y^2}}^\text{Ni/Cu}$ in $S$ (eV)		& $-0.46$ & $-0.46$ &  $-0.46$ & $-0.34$ & $-0.44$ \\
		$\epsilon_{3d_{x^2-y^2}}^\text{Ni/Cu}$ in $S-3$ (eV)	& $-1.22$ & $-1.24$ & $-1.27$ & $-1.93$ & $-1.85$ \\
		$\Delta \epsilon_{3d_{x^2-y^2}}^\text{Ni/Cu}$ (eV)		& $0.76$  & $0.78 $ & $0.81$  & $1.59$  & $1.41$  \\
		$\Delta \epsilon_{3d_{x^2-y^2}}^\text{Ni/Cu} / d_{\text{Film}}$ (meV$/\AA$)	& $56$ & $57$ & $58$ & $118$ & $96$ \\
		\hline
		$\epsilon_{3d_{xy}}^\text{Ti}$ in $S-4$ (eV)			& $-0.54$ & $-0.50$ & $-0.41$ & $-0.21$ & $-0.28$ \\
		\hline
		$\epsilon_{5d~\text{hybrid state}}$ in the bulk (eV)		& $-0.48$ & $-0.42$ & $-0.33$ & ---     & ---     \\
		$\epsilon_{5d~\text{hybrid state}}$ at the interface (eV)	& $-0.63$ & $-0.69$ & $-0.81$ & ---     & ---     \\
	\end{tabular}
	\end{ruledtabular}
\end{table*}

\begin{figure}[b]
	\centering
	\includegraphics[width=8.3cm]{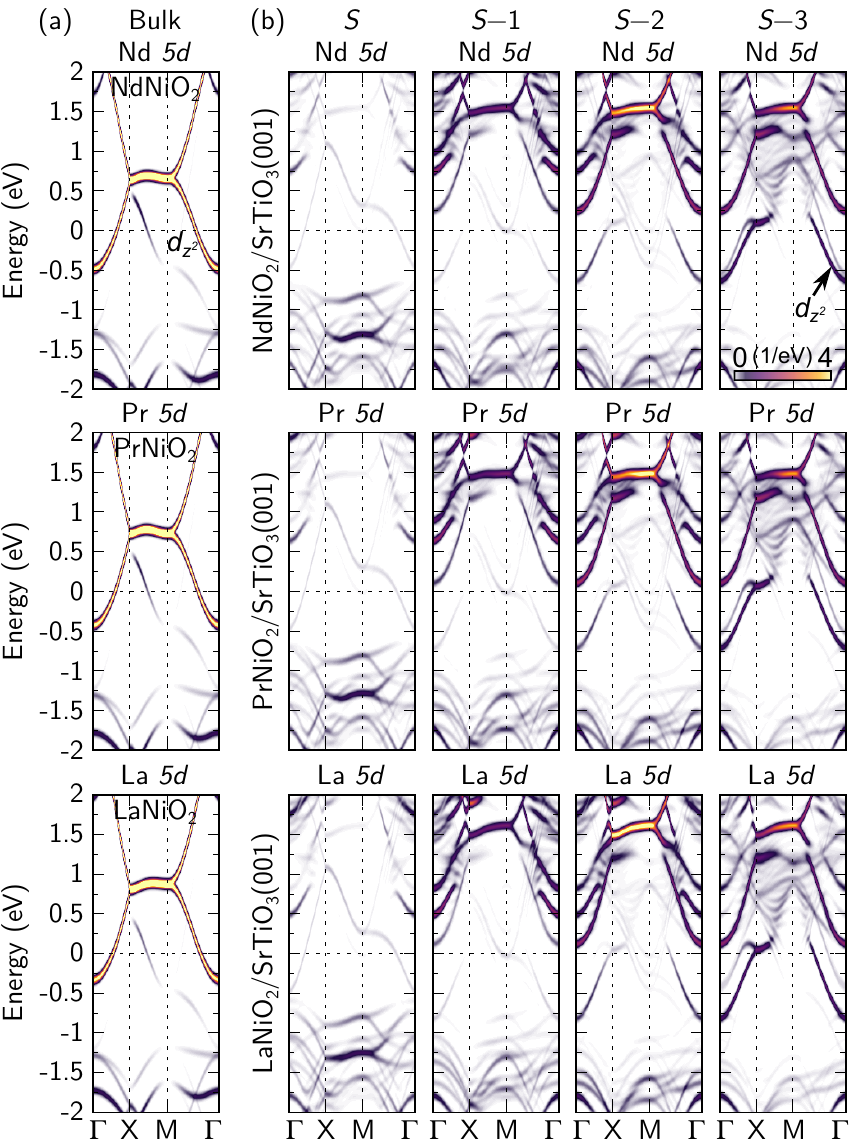}
	\caption{(a)~The rare-earth $5d$ states, which hole-dope the NiO$_2$ layers in bulk NdNiO$_2$, PrNiO$_2$, and LaNiO$_2$ due to electron pockets (for instance, at the $\Gamma$ point), (b)~contribute to the Fermi surface of NdNiO$_2$/STO$(001)$ \cite{GeislerPentcheva-InfNNO:20}, PrNiO$_2$/STO$(001)$, and LaNiO$_2$/STO$(001)$ only directly at the interface in the form of a hybrid state. In the cuprates, the respective $5d$ states are located at higher energies~\cite{Lechermann-Inf:20}.}
	\label{fig:NdLa5d}
\end{figure}

\section{Electronic reconstruction: Correlated \lc{q}2DEG formation and cuprate-like Fermi surfaces}

We now explore the implications of the polar discontinuities at the interface and the surface on the electronic structure.
As already suggested by the ionic relaxations near the interface [Fig.~\ref{fig:Structures}(c)],
we find q2DEGs to emerge in the substrate for each case (Fig.~\ref{fig:LDOS-ILDOS}),
as reported earlier for NdNiO$_2$/STO$(001)$~\cite{GeislerPentcheva-InfNNO:20}.
All four ($AB$O$_2$)$_4$/STO$(001)$ systems show a strong Ti $3d$ occupation at the interface,
in stark contrast with the paradigmatic (LAO)$_4$/STO(001) system which is just at the verge of a metal-insulator transition~\cite{PentchevaPickett:09}.
Near the interface, each q2DEG is formed predominantly by dispersive Ti $3d_{xy}$ states,
as observable in the layer-resolved band structures compiled in Fig.~\ref{fig:Bands}.
This goes hand in hand with the
ferroelectric-like Ti displacements [Figs.~\ref{fig:Structures}(c,d)]
and resembles the situation in LAO/STO(001)~\cite{PentchevaPickett:08, PentchevaPRL:10}.
The orbital order, which is also visible in the distribution of the electron density (Fig.~\ref{fig:LDOS-ILDOS}), persists within the topmost three layers and then develops into a uniform occupation of the $t_{2g}$ manifold.
Notably, the infinite-layer compounds exhibit a quite covalent nature in the $B$O$_2$ layers,
as reflected in the distribution of the electron density (Fig.~\ref{fig:LDOS-ILDOS}),
while the Ti states in the substrate are far more localized.

Surprisingly, the q2DEG formation in STO$(001)$ occurs despite the metallic character of the films
that could screen the polarity mismatch at the interface.
Specifically, for a reduced polarity mismatch at the interface,
as present for instance in metallic NdNiO$_3$/STO$(001)$ with a (NdO)$^{1+}/$(TiO$_2$)$^{0}$ interface stacking,
no q2DEG forms~\cite{GeislerPentcheva-InfNNO:20}.

We find that the film composition tunes the manifestation of the q2DEG, reflected in the different band bending (local electrostatic modification of the energy eigenvalues) of the Ti $3d_{xy}$ orbital,
which we quantify in Table~\ref{tab:BandBending}.
It is largest for NdNiO$_2$~\cite{GeislerPentcheva-InfNNO:20} and PrNiO$_2$ films ($-0.54$ and $-0.50$~eV) and half as strong for SrCuO$_2$ and CaCuO$_2$ ($-0.28$ and $-0.21$~eV),
which is in line with the lower formal polarity mismatch in the cuprate case.
The distinct band bending of the STO valence states observable in Fig.~\ref{fig:LDOS-ILDOS}
indicates different band offsets of the present infinite-layer nickelates vs.\ cuprates.

\begin{figure*}
	\centering
	\includegraphics[]{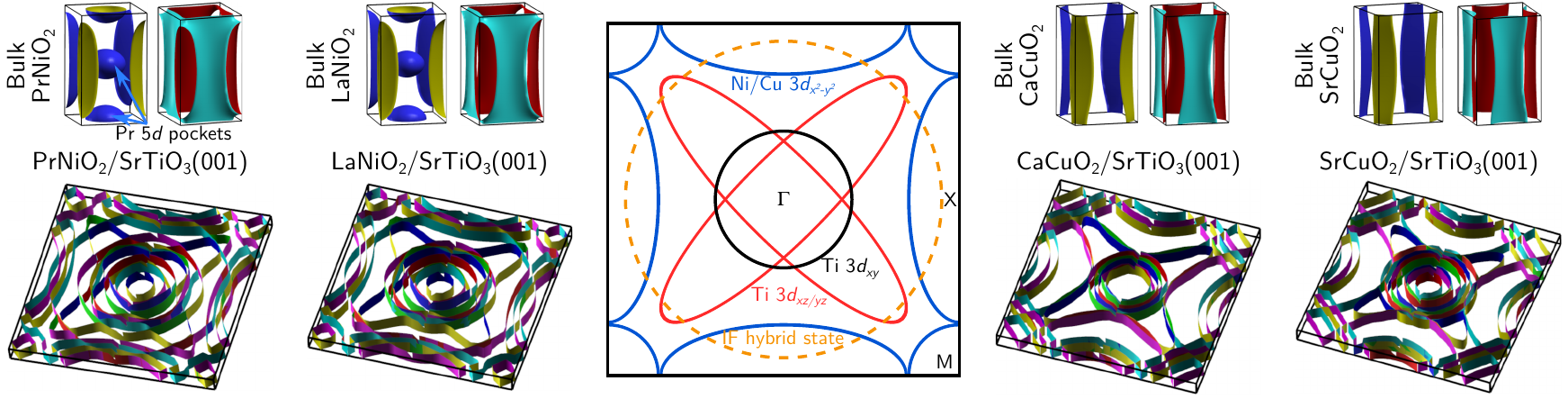}
	\caption{Fermi surfaces of $AB$O$_2$/STO$(001)$ and the corresponding infinite-layer bulk compounds in comparable $\sqrt{2} \times \sqrt{2}$ cells. Particularly for the nickelate films, the Fermi surfaces are strongly reconstructed with respect to the bulk, owing to the depletion of the rare-earth $5d$ states visible as pockets around the $\Gamma$ and $Z$ points (cf.~Fig.~\ref{fig:NdLa5d}), which considerably enhances the similarity to cuprates. The remaining difference consists predominantly of the hybrid interface state. The illustration in the center disentangles the distinct contributions to the complex Fermi surface. The repetition of sheets associated with the Ti $t_{2g}$ or the Ni/Cu $3d_{x^2-y^2}$ states reflects the electrostatic modulation.}
	\label{fig:FermiSurfaces}
\end{figure*}

It is so far unresolved why NdNiO$_2$/STO$(001)$ \cite{Li-Supercond-Inf-NNO-STO:19, Li-Supercond-Dome-Inf-NNO-STO:20}
and PrNiO$_2$/STO$(001)$ \cite{Osada-PrNiO2-SC:20}
exhibit superconductivity,
whereas such a phase is absent in LaNiO$_2$/STO$(001)$ \cite{Li-Supercond-Inf-NNO-STO:19}
despite a similar electronic structure of all three infinite-layer nickelates in the bulk.
Magnetic interactions with the rare-earth $5d/4f$ electrons have been suggested to possibly enter the mechanism~\cite{Sawatzky-NNO:19, Choi-Lee-Pickett-4fNNO:20}.
However, superconductivity could not be confirmed experimentally in Sr-doped bulk NdNiO$_2$~\cite{Li-NoSCinBulkDopedNNO:19},
which raised a question about the role of the interface and the film geometry~\cite{GeislerPentcheva-InfNNO:20}.
If we speculate that superconductivity
is mediated by the q2DEG,
as it is the case in LAO/STO$(001)$~\cite{Reyren:07},
this would require notable differences in the q2DEG for NdNiO$_2$ and PrNiO$_2$ vs.\ LaNiO$_2$.
Indeed, we observe that the Ti $3d_{xy}$ band bending is about $20$-$30\%$ larger for PrNiO$_2$ and NdNiO$_2$ than for LaNiO$_2$ (Table~\ref{tab:BandBending}, Fig.~\ref{fig:Bands}).
The distinct carrier concentration implied by these differences in electrostatic doping could drive the system out of the superconducting dome~\cite{Li-Supercond-Dome-Inf-NNO-STO:20}.
Further possible reasons for the absence of superconductivity in LaNiO$_2$/STO$(001)$
are an inhibited q2DEG formation owing to incomplete reduction of the initial perovskite nickelate films in experiment~\cite{Li-Supercond-Inf-NNO-STO:19},
as predicted for NdNiO$_2$/STO$(001)$~\cite{GeislerPentcheva-InfNNO:20},
or hydrogen intercalation following the topotactic reduction reaction~\cite{Si-Zhonh-Held:InfNNO-Hydrogen:20}.

In bulk infinite-layer cuprates, a single hole occupies the planar Cu $3d_{x^2-y^2}$ orbital~\cite{Botana-Inf-Nickelates:19, Lechermann-Inf:20}.
The situation is modified in bulk nickelates due to two self-doping electron pockets (at the $\Gamma$ and the $A$ point, the latter corresponding to the $Z$ point in the present geometry) that exhibit rare-earth $5d$ character [Figs.~\ref{fig:Bands} and~\ref{fig:NdLa5d}(a)]~\cite{Nomura-Inf-NNO:19, JiangZhong-InfNickelates:19, Sakakibara:19, JiangBerciuSawatzky:19, Botana-Inf-Nickelates:19, Choi-Lee-Pickett-4fNNO:20, Lechermann-Inf:20}.
In film geometry, the Ni and Cu $3d_{z^2}$ orbital remains completely occupied,
while electrostatic doping induces a layer-wise modulation of the $3d_{x^2-y^2}$ orbital occupation (Fig.~\ref{fig:Bands}; cf.~Fig.~\ref{fig:OrbitalProjections} in the Appendix).
This is reflected in the band energies $\epsilon$ shown in Table~\ref{tab:BandBending}.
The magnitude of the modulation throughout the film is expressed by the difference of these band energies $\Delta \epsilon$.
Counterintuitively, it turns out to be approximately twice as strong in the cuprate films ($1.41$, $1.59$~eV) than in the nickelate films ($0.76$, $0.78$, $0.81$~eV), even if normalized to the film thickness and
despite the higher polarity mismatch at the surface and the interface in the nickelate case (Table~\ref{tab:BandBending}, Figs.~\ref{fig:LDOS-ILDOS} and~\ref{fig:Bands}).

The involvement of the Nd $5d$ states
in the Fermi surface and the superconductivity mechanism of NdNiO$_2$ is currently intensely discussed~\cite{Sawatzky-NNO:19, NNO-SelfDopingDesign-d9-Arita:20, NNO-SC-Thomale:20}.
In the bulk, the rare-earth $5d$ electron pocket around the $\Gamma$ point is smaller for LaNiO$_2$ than for PrNiO$_2$ and NdNiO$_2$,
and the states extend down to $-0.33$~eV ($-0.42$ and $-0.48$~eV) below the Fermi energy in the former (latter) case [Table~\ref{tab:BandBending}, Fig.~\ref{fig:NdLa5d}(a)].
In film geometry, however, we find the rare-earth $5d$ states to be almost entirely depleted in all nickelate systems [Fig.~\ref{fig:NdLa5d}(b)].
Exclusively at the interface, they hybridize with Ni and particularly Ti $3d_{z^2}$ orbitals,
which enhances their band bending below the Fermi energy.
Surprisingly, this effect is stronger for LaNiO$_2$ films (bent down to $-0.81$~eV) than for PrNiO$_2$ and NdNiO$_2$ films (bent down to $-0.69$ and $-0.63$~eV; Table~\ref{tab:BandBending}).
We speculate that the resulting interface state partially compensates the polar discontinuity at the interface for the nickelate films
and thereby reduces the electrostatic modulation $\Delta \epsilon$ experienced by the planar $3d_{x^2-y^2}$ orbitals relative to the cuprate films (Table~\ref{tab:BandBending}).

The Fermi surfaces of the different $AB$O$_2$/STO$(001)$ systems shown in Fig.~\ref{fig:FermiSurfaces}
demonstrate how the electronic reconstruction enhances the similarity of nickelates and cuprates,
owing to the depletion of the two rare-earth $5d$ electron pockets in the nickelate films.
The remaining differences are largely of quantitative nature and arise due to variations in the q2DEG in the topmost STO(001) layers
and the distinct degree of modulation of the planar Ni and Cu $3d_{x^2-y^2}$ orbitals.
As illustrated in Fig.~\ref{fig:FermiSurfaces},
the four-pointed-star-shaped Fermi surface sheets centered at the $\Gamma$ point reflect the q2DEG emerging in the STO conduction band,
specifically the Ti $3d_{xz/yz}$ orbitals that cross the Fermi energy at a few layers distance to the interface,
whereas the Ti $3d_{xy}$ orbitals give rise to circular sheets that are also centered at the $\Gamma$ point~\cite{STO-HimmetogluJanottiWalle:14, STO-FS-Chen:15}.
The splitting of the Ti $3d_{xy}$ and $3d_{xz/yz}$ orbitals is induced by the electrostatic doping and ionic relaxations at the interface.
In the case of rare-earth nickelate films, the $d_{z^2}$ hybrid interface state is represented by a large circular sheet centered at the $\Gamma$ point,
whereas the $5d$ electron pockets are empty and thus absent.
The features at the Brillouin zone boundary are contributed by the planar Ni and Cu $3d_{x^2-y^2}$ orbitals.

\begin{figure}
	\centering
	\includegraphics[]{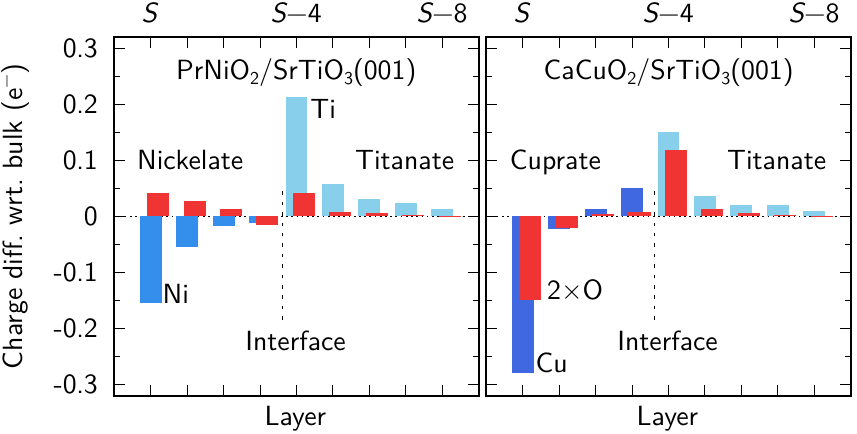}
	\caption{Layer-resolved charge difference in the representative systems PrNiO$_2$/STO$(001)$ vs.\ CaCuO$_2$/STO$(001)$ relative to the constituent bulk compounds, integrated at the Ni and Cu sites (dark blue), Ti sites (light blue), and in the corresponding basal oxygen sublattice (red). The highly distinct response of the latter in nickelates vs.\ cuprates reveals an intra-layer charge transfer in the nickelates in addition to the common charge transfer from the surface to the interface.}
	\label{fig:ChargeDifferences}
\end{figure}

We complete the discussion of the polarity-driven electronic reconstruction
by contrasting the layer- and site-resolved charge differences as they arise for infinite-layer nickelates vs.\ cuprates
in film geometry
with respect to the corresponding bulk systems (Fig.~\ref{fig:ChargeDifferences}).
Both representative systems PrNiO$_2$/STO$(001)$ and CaCuO$_2$/STO$(001)$ show a depletion of electrons near the surface
and a concomitant accumulation in the interfacial Ti layers,
which constitutes the correlated q2DEG and rapidly decays into the substrate.
The decay is paralleled by the decreasing ferroelectric-like displacements of the Ti ions reported above [Fig.~\ref{fig:Structures}(c)].
While the Ni sites show a loss of electrons throughout the film,
the Cu sites exhibit a loss exclusively near the surface and a slight gain near the interface.
Interestingly, the oxygen sublattice responds highly differently to the polar discontinuity in nickelate vs.\ cuprate films:
In the nickelate case, the oxygen sites largely gain charge as opposed to the Ni sites,
i.e., some electrons are transferred from Ni to oxygen within each NiO$_2$ layer.
In the cuprate case, the oxygen sites parallel the behavior observed at the Cu sites,
so that charge is redistributed exclusively between $B$O$_2$ layers.
This highlights the different degree of $B$~$3d$-O~$2p$ hybridization in the two materials classes.
In the substrate, the oxygen sites always gain charge, particularly for CaCuO$_2$.
The loss at the surface is roughly twice as strong for CaCuO$_2$ as for PrNiO$_2$,
consistent with the much larger electrostatic modulation (Table~\ref{tab:BandBending}, Figs.~\ref{fig:LDOS-ILDOS} and~\ref{fig:Bands}).
Hence, we conclude that the charge redistribution in the systems is not simply proportional
to the formal polarity mismatch between the infinite-layer film and the nonpolar substrate,
but unravels a complex interplay of the emergent interface electronic structure and the screening characteristics of the film.

\begin{figure}
	\centering
	\includegraphics[width=6.0cm]{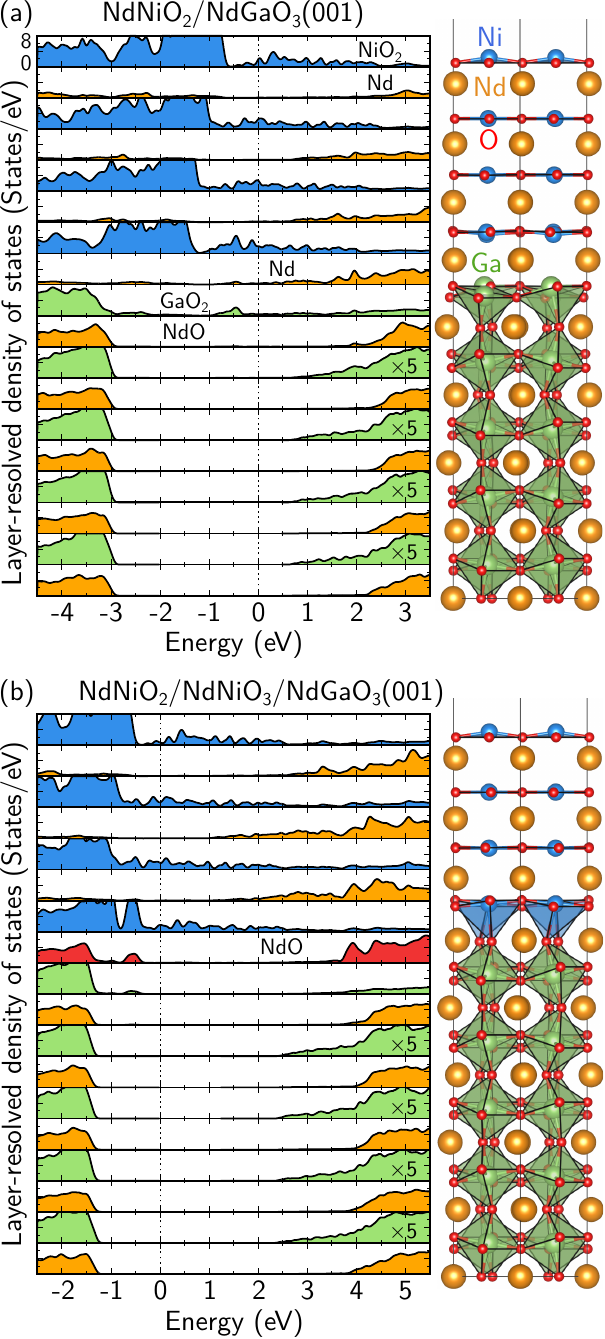}
	\caption{(a)~In NdNiO$_2$/NdGaO$_3(001)$, the completely filled Ga~$3d$ shell of the substrate leads to physics fundamentally different from NdNiO$_2$/SrTiO$_3(001)$ \cite{GeislerPentcheva-InfNNO:20}, characterized by the absence of the interfacial q2DEG, but retaining the depleted Nd $5d$ states in the infinite-layer film. In the layer-resolved density of states, the GaO$_2$ conduction band states have been enhanced for better visibility. (b)~In the case of an oxidized interface layer, the band alignment changes from $n$- to $p$-type, at variance with NdNiO$_2$/SrTiO$_3(001)$ which is always $n$-type~\cite{GeislerPentcheva-InfNNO:20}.}
	\label{fig:NGO}
\end{figure}

\section{\boldmath Role of the substrate: N\lc{d}N\lc{i}O$_2$/N\lc{d}G\lc{a}O$_3$(001)}

While most experiments on superconducting infinite-layer nickelates have been conducted on nonpolar STO$(001)$ so far,
further insight into the superconductivity mechanism could be gained by exchanging STO with a typical insulating substrate with
naturally alternating formal charge of the consecutive pseudocubic $(001)$ layers,
for instance, LaGaO$_3$ (pseudocubic lattice constant: $3.90~\AA$) or NdGaO$_3$ ($3.86~\AA$).
The degree of compressive strain induced by these substrates is comparable to that exerted by STO ($3.905~\AA$),
whereas the lattice constants of typical aluminates such as LAO ($3.79~\AA$) are considerably smaller.
Exemplarily, Fig.~\ref{fig:NGO}(a) shows the optimized geometry
and layer-resolved electronic structure of NdNiO$_2$/NdGaO$_3(001)$.
In this system, the interface Nd$^{3+}$ layer is closer in formal charge to the (NdO)$^{1+}$ layers in the substrate,
in contrast to the (SrO)$^{0}$ layers in a STO substrate.
Hence, while the formal polarity mismatch at the Nd$^{3+}$/(GaO$_2$)$^{1-}$ interface is even higher than at the STO$(001)$ interface,
the infinite-layer film shows a comparable electrostatic modulation.
In contrast to STO, the NdGaO$_3$ substrate exhibits strong $a^-a^-c^+$ octahedral rotations
that induce modest $c^-$ rotations the NiO$_2$ plaquettes near the interface.
NiO$_2$ buckling is observed exclusively at the surface, at variance with the nickelate films grown on the STO$(001)$ substrate (cf.~Fig.~\ref{fig:Structures}).
The completely filled Ga~$3d$ shell of the substrate leads to fundamentally different behavior from NdNiO$_2$/SrTiO$_3(001)$ \cite{GeislerPentcheva-InfNNO:20},
characterized by a quenched q2DEG, but retaining the depleted Nd $5d$ states, i.e.,
a cuprate-like electronic structure in the nickelate film.
The Fermi energy is located $\sim 0.7$~eV below the conduction band of NdGaO$_3$, resulting in an $n$-type electronic structure.
In the case of an oxidized interface layer [formally (NdNiO$_2$)$_3$/(NdNiO$_3$)$_1$/NdGaO$_3(001)$, Fig.~\ref{fig:NGO}(b)], the band alignment changes from $n$- to $p$-type,
and the Fermi energy is now located $\sim 1.2$~eV above the valence band of NdGaO$_3$.
This situation is in sharp contrast to NdNiO$_2$/SrTiO$_3(001)$ with oxidized interface layer, which is $n$-type~\cite{GeislerPentcheva-InfNNO:20}.
The oxygen vacancy formation energy 
$E_f^{} = E_{\text{Nd/GaO$_2$ interface}}^{} - E_{\text{NdO/GaO$_2$  interface}}^{} + \frac{1}{2} E_\text{O$_2$}$
(oxygen-rich limit)
amounts to $3.8$~eV and is hence lower than in NdNiO$_2$/SrTiO$_3(001)$ ($4.1$~eV~\cite{GeislerPentcheva-InfNNO:20}), which may facilitate a complete reduction of the nickelate film during the topotactic reaction.

Table~\ref{tab:Summary-2DEG} compiles results for the q2DEG formation in different perovskite and infinite-layer systems (at 4 ML film thickness)
as a function of the formal polarity mismatch at the interface
and puts them into an interesting context.
In the paradigmatic band insulator system LAO/STO$(001)$, a polarity mismatch of ${1+}/0$ is sufficient to drive the emergence of a q2DEG.
In contrast, NdNiO$_3$/STO$(001)$ does not develop a q2DEG despite having equal interface polarity, owing to metallic screening in the film.
The further increased interface polarity in the infinite-layer cuprates (${2+}/0$) and nickelates (${3+}/0$) on STO$(001)$ leads to the emergence of a very pronounced q2DEG despite the metallic screening present in particular in the nickelates.
Replacing the nonpolar STO substrate with polar NdGaO$_3$ quenches the q2DEG entirely due to the closed Ga $3d$ shell.

\begin{table}
	\centering
	\caption{\label{tab:Summary-2DEG}Emergence of interfacial q2DEGs for different perovskite and infinite-layer films on STO$(001)$ and NdGaO$_3(001)$ as a function of the formal polarity mismatch at the interface.}
	\begin{ruledtabular}
	\begin{tabular}{lcc}
		System & Interface polarity & q2DEG \\
		\hline
		LaAlO$_3$/SrTiO$_3(001)$ \cite{Thiel:06, PentchevaPickett:09} & ${1+}/0$ & yes \\
		NdNiO$_3$/SrTiO$_3(001)$ \cite{GeislerPentcheva-InfNNO:20} & ${1+}/0$ & no \\
		CaCuO$_2$/SrTiO$_3(001)$ & ${2+}/0$ & yes \\
		SrCuO$_2$/SrTiO$_3(001)$ & ${2+}/0$ & yes \\
		NdNiO$_2$/SrTiO$_3(001)$ \cite{GeislerPentcheva-InfNNO:20} & ${3+}/0$ & yes \\
		PrNiO$_2$/SrTiO$_3(001)$ & ${3+}/0$ & yes \\
		LaNiO$_2$/SrTiO$_3(001)$ & ${3+}/0$ & yes \\
		NdNiO$_2$/NdGaO$_3(001)$ & ${3+}/{1-}$ & no \\
	\end{tabular}
	\end{ruledtabular}
\end{table}

\section{Summary}

The impact of interface polarity on the structural and electronic properties of PrNiO$_2$/SrTiO$_3$(001), LaNiO$_2$/SrTiO$_3$(001), CaCuO$_2$/SrTiO$_3$(001), and SrCuO$_2$/SrTiO$_3$(001) was investigated by performing first-principles calculations in film geometry including a Coulomb repulsion term.
Similar to NdNiO$_2$/SrTiO$_3$(001),
polar discontinuity drives the emergence of a quasi-two-dimensional electron gas (q2DEG) at the interface in all cases
due to the occupation of the Ti $3d$ conduction band
that is accompanied by substantial ferroelectric-like displacements of the Ti ions.
%
Despite their comparable electronic structure in the bulk,
the higher polarity mismatch at the interface of infinite-layer nickelates vs.\ cuprates to the nonpolar SrTiO$_3(001)$ substrate
enhances the q2DEG carrier density for the nickelate films.
%
In addition, we found a strong dependence of the carrier density on the rare-earth ion in the nickelate films, being larger for PrNiO$_2$ and NdNiO$_2$ than for LaNiO$_2$.
This difference in carrier density could affect the superconducting properties of the q2DEG itself.
%
On the other hand,
the depletion of the self-doping rare-earth $5d$ states enhances the similarity of nickelate and cuprate Fermi surfaces in film geometry,
except for a $5d$-$3d$ hybrid interface state present for nickelates.
The resulting single hole in the Ni and Cu $3d_{x^2-y^2}$ orbitals is modulated throughout the infinite-layer films due to electrostatic doping, which turns out to be twice as strong in cuprates as in nickelates, contrary to expectations from the formal polarity mismatch.
These results highlight similarities, but also fundamental differences between infinite-layer nickelates and cuprates,
and provide clues as to why nickelate superconductivity is so far exclusively observed in film geometry.
Finally, we explored NdNiO$_2$ films grown on a polar NdGaO$_3(001)$ substrate, and showed that no q2DEG emerges at the interface, while simultaneously the Nd~$5d$ states in the film are depleted.
This promotes NdGaO$_3(001)$ as interesting substrate that may provide deeper insight into the superconductivity mechanism in infinite-layer nickelates.

\section{Acknowledgments}


\begin{figure*}
	\centering
	\includegraphics[width=13cm]{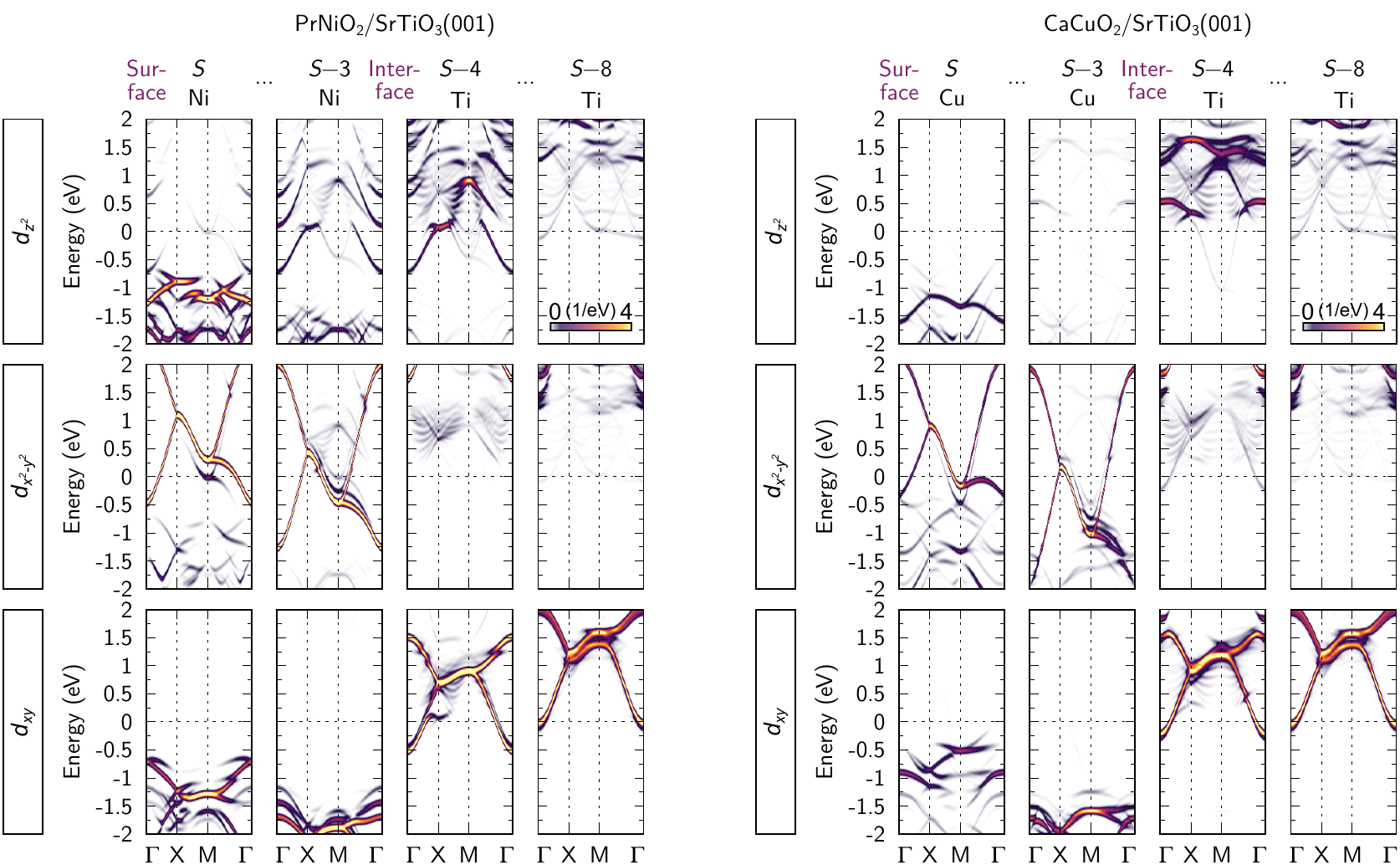}
	\caption{Orbital- and layer-resolved band structures (cf.~Fig.~\ref{fig:Bands}) of the representative systems PrNiO$_2$/STO$(001)$ and CaCuO$_2$/STO$(001)$. Projections on Ni, Cu, and Ti sites in selected layers near the surface, the interface, and deep within the STO$(001)$ substrate are shown. }
	\label{fig:OrbitalProjections}
\end{figure*}

This work was supported by the German Research Foundation (Deutsche Forschungsgemeinschaft, DFG) within the SFB/TRR~80 (Projektnummer 107745057), Project No.~G3.
Computing time was granted by the Center for Computational Sciences and Simulation of the University of Duisburg-Essen
(DFG Grants No.~INST 20876/209-1 FUGG and No.~INST 20876/243-1 FUGG)
and by the Leibniz-Rechenzentrum, Garching bei München (Grant No.~pr87ro).


\appendix*

\section{Orbital contributions to the electronic structure}

In order to disentangle the different orbital contributions to the electronic structure shown in Fig.~\ref{fig:Bands},
Fig.~\ref{fig:OrbitalProjections} displays a selection of orbital- and layer-resolved band structures
for the representative systems PrNiO$_2$/STO$(001)$ and CaCuO$_2$/STO$(001)$.
Similar to the bulk, the Ni and Cu $3d_{z^2}$ and $t_{2g}$ states (exemplarily, the $3d_{xy}$ orbitals are shown) are completely occupied in both nickelate and cuprate films.
The resulting single hole in the Ni and Cu $3d_{x^2-y^2}$ orbitals is modulated throughout the infinite-layer films due to electrostatic doping.
The hybrid interface state formed by Ni $3d_{z^2}$, Ti $3d_{z^2}$, and the rare-earth $5d_{z^2}$ states appears exclusively for nickelate films.
At the interface, the emerging q2DEG is constituted predominantly by Ti $3d_{xy}$ states.


%

\end{document}